\documentstyle[12pt]{article}
\begin{document}

\begin{titlepage}
\begin{flushright}
{\tt hep-ph/9902290}
\end{flushright}
\vspace{2cm}

\begin{center}
{\large\bf  Infrared Quasi Fixed Points  and Mass Predictions in the
MSSM II: Large $\tan\beta$ Scenario} \\[1cm]

  {\bf  M. ~Jur\v{c}i\v{s}in\footnote{
On leave of absence from the Institute of Experimental Physics, SAS,
Ko\v{s}ice, Slovakia}  and  D. I.~Kazakov}\\[5mm]

{\it Bogoliubov Laboratory of Theoretical Physics,
Joint Institute for Nuclear Research, \\
141 980 Dubna, Moscow Region, Russian Federation}

\end{center}

\vspace{2cm}

\abstract{ We consider the infrared quasi fixed point solutions of
the renormalization group equations for the  Yukawa couplings and
soft supersymmetry breaking parameters in the MSSM in the
\underline{large $\tan\beta$} regime. The existence of IR quasi
fixed points together with the values of  gauge couplings, third
generation quarks, lepton  and Z-boson masses allows one to
predict  masses of the  Higgs bosons and SUSY particles as
functions of the only free parameter, $m_{1/2}$, or the gluino
mass.  The lightest Higgs boson mass for  $M_{SUSY} \approx 1$ TeV
is found to be $m_h=128.2-0.4-7.1 \pm 5$ GeV for $\mu>0$ and
$m_h=120.6-0.1-3.8 \pm 5$ GeV  for $\mu<0$.}
\end{titlepage}

\section{Introduction}

Supersymmetric extensions of the Standard Model (SM) are believed
to be the most promising theories at high energies. An attractive
feature of SUSY theories is a possibility of unifying various
forces of Nature.  The best known supersymmetric extension of the
SM is the so-called  Minimal Supersymmetric Standard Model (MSSM)
\cite{1}. The parameter freedom of the MSSM comes mainly from the
so-called soft SUSY breaking terms, which are the sources of
uncertainty in the MSSM predictions. The most common way to reduce
this uncertainty is to assume universality of  soft terms, which
means an equality of some parameters at a high energy scale.
Adopting the universality, one reduces the parameter space to a
five-dimensional one \cite{1}: $m_0$,  $m_{1/2}$, $A$, $\mu$, and
$B$. The last two parameters it is convenient to trade for the
electroweak scale $v^2=v_{1}^2+v_{2}^2=(174.1 GeV)^2$, and
$\tan\beta=v_{2}/v_{1}$, where $v_{1}$ and $v_{2}$ are the Higgs
field vacuum expectation values. In  papers \cite{2}-\cite{6}
further reduction of the parameter space has been  discussed based
on the concept of the so-called infrared quasi fixed points
(IRQFP)~\cite{7,101}, which shows insensitivity of some low-energy
MSSM parameters to their initial high energy values. This opens a
possibility to compute physical values of some masses and soft
couplings at low-energies without detailed knowledge of physics at
high energies. For example, in Ref.~\cite{2b} the dependence of
sparticle spectrum on two free parameters, namely $m_0$ and
$m{1/2}$, has been discussed. In paper \cite{6} we argue that for
some sparticles and Higgs bosons it is possible to reduce this
two-dimensional parameter space into one-dimensional one using the
IRQFPs. The only relevant parameter is $m_{1/2}$ which is directly
related to gluino mass $M_3$.

In the literature the IRQFP scenario has been considered in detail
mainly for the MSSM in the low $\tan\beta$ regime.
In the present paper we make an analysis similar to that in
Ref.\cite{6} but for  large $\tan\beta$. In this case the
reduction of the parameter space is also possible and the
IRQFPs  also appear. The situation, however,  is more complicated
because now we have the whole set of Yukawa couplings of the third
generation  and the corresponding soft trilinear terms. Here
no exact solution of the renormalization group equations (RGEs) is
known and one is bound to a numerical investigation.  Nevertheless, we
show that the IRQFP are well pronounced and may be used to calculate
a mass spectrum.

In what follows we assume the equality of all three Yukawa
couplings at the GUT scale, i.e.  we work in the so-called
$SO(10)$ unification scheme which is  natural  for large
$\tan\beta$. The other possibility, $h_t \approx h_b > h_{\tau}$,
is not considered here.  We use the obtained IRQFPs to make
predictions for the Higgs masses and masses of the stops and
sbottoms as functions of the only free parameter, namely the
gluino mass, $M_3$. We also present  the mass of the lightest
Higgs boson as a function of the geometric mean of stop masses,
sometimes called  $M_{SUSY}$, and confront our results with the
experimental data on  SUSY and Higgs boson searches.

The paper is organized as follows. In Section 2 we present the
numerical analysis of  the RGE's and demonstrate their infrared
behaviour. IRQFPs are found and used in Section 3 to obtain the
mass spectrum of  Higgs bosons and some SUSY particles. In Section
4 we discuss our main results and conclusions.

\section{Infrared Quasi Fixed Points and RGE's}

We now give a short description of the infrared behaviour of the
RGE's in the MSSM for the large $\tan\beta$ regime. We follow the
same strategy as in our previous paper \cite{6} related to the low
$\tan\beta$ case, though  the analysis
for large $\tan\beta$ is more complicated because one has
 to take into account three Yukawa couplings,
$Y_{i}$, where $Y_{i}=h_i^2/(4\pi)^2$ and $i=(t, b, \tau)$ ($t$
corresponds to top-quark, $b$ to bottom-quark and $\tau$ to
$\tau$-lepton) and  the corresponding three trilinear SUSY
breaking parameters $A_i$. Since the analytical solution of the
RGE's is unknown (some semianalytical solutions for appropriate
combinations of the parameters exist~\cite{9} but they are almost
irrelevant for our analysis), we  use  numerical methods.

Recently it has been proven~\cite{10} that in the asymptotically
free case the existence of stable IR fixed points for the Yukawa
couplings implies stable IR fixed points for A-parameters and soft
scalar masses. (It follows also from superfield description of
softly broken SUSY theories~\cite{D}.) Thus, though we are not
able to make an analytical analysis as has been done in
Ref.\cite{6}, we can solve RGE's numerically and find infrared
fixed points. Here the same concept of IRQFP arises and one can
find the intervals in which initial parameters can vary to be
attracted by these fixed points.

We begin with an analysis of  RGE's for the Yukawa couplings. It
is a self-consistent system of differential equations (together
with well-known RGE's for gauge couplings), which in the one-loop
order has the following form (see e.g. Ref.\cite{K}):
\begin{eqnarray}
\frac{d Y_t}{d t}
&=& Y_t \left( \frac{16}{3} \tilde\alpha_3 + 3 \tilde\alpha_2
+ \frac{13}{15} \tilde\alpha_1 - 6 Y_t - Y_b\right)\,, \label{Yt}   \\
\frac{d Y_b}{d t}
&=& Y_b \left( \frac{16}{3} \tilde\alpha_3 + 3 \tilde\alpha_2
+ \frac{7}{15} \tilde\alpha_1 -  Y_t - 6 Y_b - Y_{\tau}\right)\,,
\label{Yb} \\
\frac{d Y_{\tau}}{d t}
&=& Y_{\tau} \left( 3 \tilde\alpha_2 + \frac{9}{5} \tilde\alpha_1
- 3 Y_b - 4 Y_{\tau}\right)\,, \label{Ytau}   
\end{eqnarray}
where $\tilde  \alpha_i= \alpha_i/(4\pi)$,
$t=\log(M_{GUT}^2/Q^2)$. As it has been mentioned in the
introduction we assume the equality  of the Yukawa couplings of
the third generation at the GUT scale ($M_{GUT}=10^{16}$ GeV):
$Y_t(M_{GUT})=Y_b(M_{GUT})=Y_{\tau}(M_{GUT})$.

\begin{figure}[t]
     \vspace{-11.5cm}
       \begin{flushleft}
       \leavevmode
       \epsfxsize=16cm
       \epsffile{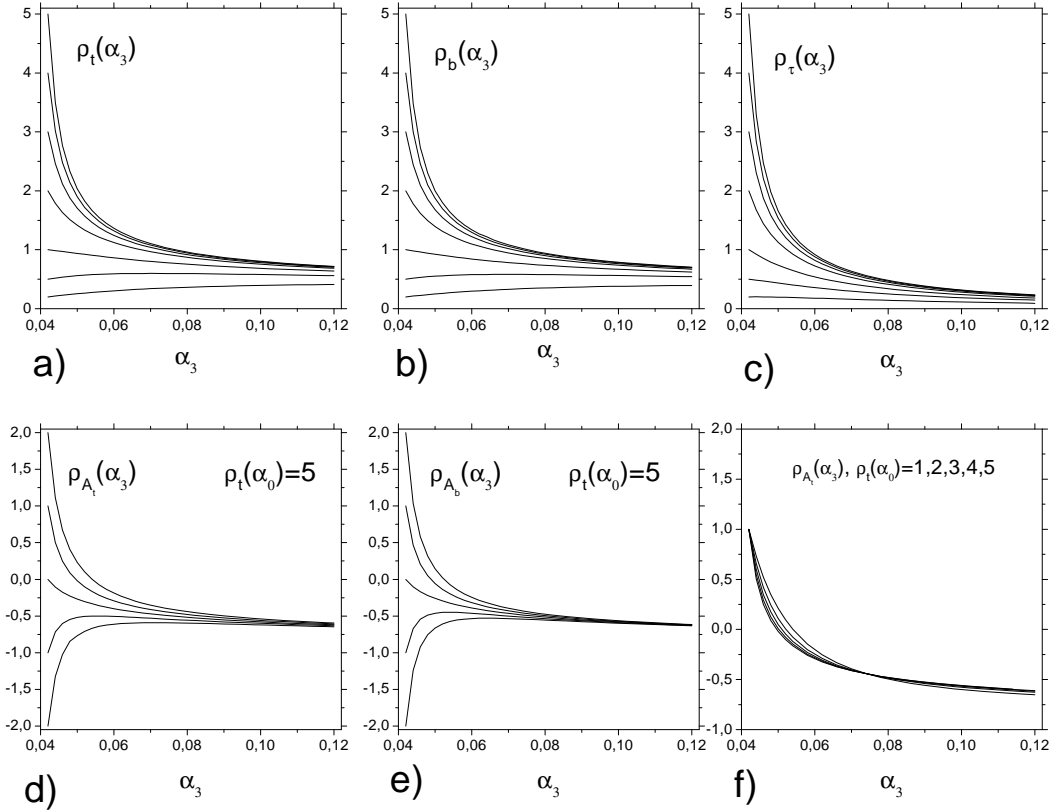}
   \end{flushleft}
\caption{The infrared quasi fixed points for $\rho_i=Y_i/
\tilde\alpha_3$ $i=t,b,\tau$ (a,b,c), $\rho_{A_i}=A_i/ M_3$
$i=t,b$ (d,e) and $\rho_{A_t}$ with $\rho_{A_t}(\alpha_0)=1$ for
different initial values of $\rho_t(\alpha_0)$ (f). \label{f1}}
\end{figure}

In Figs.\,\ref{f1}a,b,c  the numerical solutions of the RGE's are
shown for a wide range of initial values of
$\rho_t(M_{GUT})=\rho_b(M_{GUT})=\rho_{\tau}(M_{GUT})$ from the
interval $<0.2, 5>$, where $\rho_i=Y_i/\tilde\alpha_3$. As one can
see from this figure, there is a strong restriction  on all the
Yukawa couplings at the scale $M_Z$. One can also clearly see the
IRQFP type  behaviour when the parameter $\rho_i$ is big enough.
Further restrictions on the initial values, which follow from the
 phenomenological arguments, are considered in the next section.

We have found the following values of the Yukawa couplings $h_i$
at the $M_Z$ scale
$$h_t \in <0.787, 1.048>, \ \ h_b \in <0.758,
0.98>,\ \ \ h_{tau} \in <0.375, 0.619>.$$
 Comparing $h_t$ and $h_b$  one can see that the ratio $h_t/h_b$
belongs to a very narrow interval $h_t/h_b \in <1.039, 1.069>$.
This property of stability helps us  to
determine $\tan\beta$ in the next section.

Now we proceed with the discussion of  RGE's for  trilinear
scalar couplings, $A_i, i=(t, b, \tau)$. The one-loop RGE's have the
following form (see e.g. Ref.\cite{K}):
\begin{eqnarray} \frac{d A_t}{d t}
&=& - \left( \frac{16}{3} \tilde\alpha_3 M_3 + 3 \tilde\alpha_2 M_2 +
\frac{13}{15} \tilde\alpha_1 M_1\right) - 6 Y_t A_t - Y_b A_b\,,
\label{At}   \\
\frac{d A_b}{d t} &=& - \left( \frac{16}{3}
\tilde\alpha_3 M_3 + 3 \tilde\alpha_2 M_2 + \frac{7}{15} \tilde\alpha_1
M_1\right) - 6 Y_b A_b - Y_t A_t - Y_{\tau} A_{\tau}\,, \label{Ab}
\\
\frac{d A_{\tau}}{d t} &=& - \left( 3 \tilde\alpha_2 M_2 +
\frac{9}{5} \tilde\alpha_1 M_1 \right) - 3 Y_b A_b - 4 Y_{\tau}
A_{\tau}\,, \label{Atau}  \\ 
\frac{d M_i}{d t}&=& - b_i  \tilde\alpha_i^2 M_i\,, \label{Mi} 
\end{eqnarray}
where $M_i$ are the  gaugino masses, $b_i$ are the one-loop
$\beta$-function coefficients for the gauge couplings
$\tilde\alpha_i$ with $(b_1, b_2, b_3)=(33/5, 1, -3)$.  This
system of  RGE's together with the equations for the Yukawa
couplings (\ref{Yt}-\ref{Ytau}) is self-consistent, and we  solve
it numerically.   The results  are shown in Figs.\,\ref{f1}d,e for
the following quantities $\rho_{A_i}=A_i/M_3, i=(t, b)$ for
different initial values at the GUT scale and for
$\rho_i(M_{GUT})=5$.  One can  see the strong
attraction to the fixed points.

The question of stability of these IRQFPs becomes important for
further consideration. We have analyzed their stability under the
change of the initial conditions for $\rho_i(M_{GUT})$ and have found
remarkable stability, which allows one to use them as  fixed
parameters at the $M_Z$ scale. In Fig.\,\ref{f1}f  a particular
example of stability of IRQFP for $A_t$ is shown. As a result
we have the following IRQFP values for the parameters
$\rho_{A_i}$:
 $$\rho_{A_t}\approx -0.619, \ \ \rho_{A_b}\approx
-0.658, \ \ \rho_{A_{\tau}}\approx 0.090.$$ The last value for
$A_{\tau}$ is not important for  calculation
 of masses (see the next section).

The last step in the investigation of the RGE's is the calculation
of the soft mass parameters and finding of appropriate IRQFPs. The
one-loop RGE's for the masses are (see e.g. Ref.\cite{K})
\begin{eqnarray} \frac{d m_Q^2}{d t} &=&
\left(\frac{16}{3} \tilde \alpha_3 M_3^2 +3 \tilde \alpha_2 M_2^2 +
\frac{1}{15} \tilde \alpha_1 M_1^2 \right) - Y_t \left(m_Q^2 +m_U^2 +
m_{H_2}^2 + A_t^2  \right) \nonumber \\ &-& Y_b \left(m_Q^2 + m_D^2 +
m_{H_1}^2 + A_b^2\right)\,, \label{mQ} \\
\frac{d m_U^2}{d t} &=&
\left(\frac{16}{3} \tilde \alpha_3 M_3^2 + \frac{16}{15} \tilde
\alpha_1 M_1^2 \right) - 2 Y_t \left(m_Q^2 +m_U^2 + m_{H_2}^2 + A_t^2
\right)\,, \label{mU} \\ 
\frac{d m_D^2}{d t} &=&
\left(\frac{16}{3} \tilde \alpha_3 M_3^2 + \frac{4}{15} \tilde \alpha_1
M_1^2 \right) - 2 Y_b \left(m_Q^2 +m_D^2 + m_{H_1}^2 + A_b^2
\right)\,, \label{mD} \\ 
\frac{d m_{H_1}^2}{d t} &=&  3 \left(
\tilde \alpha_2 M_2^2 + \frac{1}{5} \tilde \alpha_1 M_1^2 \right) - 3
Y_b \left(m_Q^2 +m_D^2 + m_{H_1}^2 + A_b^2  \right) \nonumber \\ &-&
Y_{\tau} \left(m_L^2 + m_E^2 + m_{H_1}^2 + A_{\tau}^2\right)\,,
\label{mH1} \\ 
\frac{d m_{H_2}^2}{d t} &=&  3 \left( \tilde \alpha_2 M_2^2
+ \frac{1}{5} \tilde \alpha_1 M_1^2 \right)
- 3 Y_t \left(m_Q^2 +m_U^2 + m_{H_2}^2 + A_t^2  \right)\,,
\label{mH2} \\ 
\frac{d m_L^2}{d t} &=&  3 \left(\tilde\alpha_2 M_2^2
+ \frac{1}{5} \tilde \alpha_1 M_1^2 \right)
- Y_{\tau} \left(m_L^2 +m_E^2 + m_{H_1}^2 + A_{\tau}^2  \right)\,,
\label{mL} \\ 
\frac{d m_E^2}{d t} &=&  \frac{12}{5} \tilde \alpha_1 M_1^2
- 2 Y_{\tau} \left(m_L^2 +m_E^2 + m_{H_1}^2 + A_{\tau}^2  \right)\,,
\label{mE}  
\end{eqnarray}
where $m_{H_1}^2$ and $m_{H_2}^2$ are the SUSY breaking masses from
the Higgs potential, $m_Q^2$, $m_U^2$ and $m_D^2$ are the squark
masses (here $Q$ refers to the third generation squark doublet,
$U$ to the stop singlet and $D$ to the sbottom singlet) and
$m_L^2$ and  $m_E^2$ are the slepton masses ($L$ refers to the third
generation doublet and $E$ to the stau singlet). We can express
them (as in the case of trilinear soft couplings $A_i$) via the common
gaugino mass $m_{1/2}$ or, equivalently, via the gluino mass
$M_3=(\tilde\alpha_3/\tilde\alpha_0) m_{1/2}$, when investigating
their IR fixed point behaviour.

\input epsf

\input epsf
   \begin{figure}[t]
     \vspace{-11.5cm}
       \begin{flushleft}
       \leavevmode
       \epsfxsize=16cm
       \epsffile{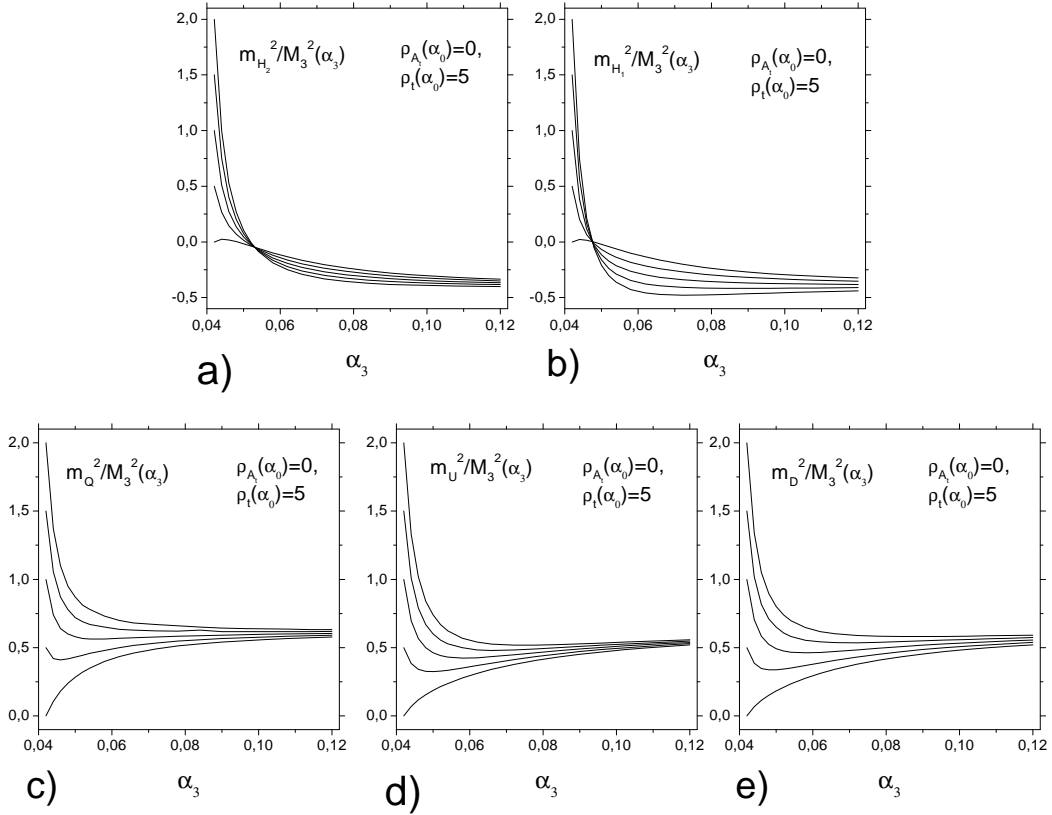}
   \end{flushleft}
\caption{The infrared quasi fixed points for mass parameters.
\label{f2}}
\end{figure}

In contrast with the low $\tan\beta$ MSSM scenario, where there is
no obvious infrared attractive fixed point for  $m_{H_1}^2$, for
large $\tan\beta$ it has almost the same infrared behaviour as
$m_{H_2}^2$ because of the non-negligible bottom-quark Yukawa
coupling in the corresponding RGE. One can see in
Figs.\,\ref{f2}a,b that both the ratios $m_{H_1}^2/M_3^2$ and
$m_{H_2}^2/M_3^2$ are negative in the infrared region. This
requires a proper value of the parameter $\mu$ for the
electro-weak symmetry breaking to take place.

As one can see from Figs. \ref{f2}a,b, there exist  IRQFP's
 $$m_{H_1}^2/M_3^2 \approx-0.306 \ \ m_{H_2}^2/M_3^2 \approx -0.339,$$
The numbers correspond to the initial condition
$m_0^2/m_{1/2}^2=0$. When calculating the mass of the Higgs bosons
one has to take into account small deviation from these fixed
points. Later on we consider  initial values for the ratio
$m_0^2/m_{1/2}^2$ belonging to the following interval
$m_0^2/m_{1/2}^2 \in <0, 2>$.

In Figs.\,\ref{f2}c,d,e the infrared behaviour of the soft SUSY
breaking squark masses is shown. One can immediately see that all
masses have IRQFPs which we use in the next section to find the
mass spectrum. For further analysis only the squark masses are
important. As for sleptons they also have an attractive infrared
behaviour but it does not influence the mass spectrum of the Higgs
bosons and we do not show them explicitly.

Numerical values of the ratios are the following:
$$m_Q^2/M_3^2 \approx 0.58, \ \ m_U^2/M_3^2 \approx 0.52, \ \
m_D^2/M_3^2 \approx 0.53,$$  obtained for $m_0^2/m_{1/2}^2=0$.  One can
refer to them as to  IRQFPs.  Small deviations from these fixed
points are again taken into account when we calculate the mass spectrum
of the Higgs bosons.
 To do this we  again address the
 question of stability of these IRQFPs. Solving the RGE's for
different initial values of the Yukawa couplings, one  again finds
a very week dependence on them.

We have analyzed also the behaviour of the bilinear SUSY breaking
parameter $B$. The  situation here is the same as in low
$\tan\beta$ case. The ratio $B/M_3$ does not exhibit the infrared
quasi fixed point behaviour and so we do not use it in further
discussion.

Thus, one can see that solutions of RGE's for all MSSM SUSY
breaking parameters (the only exception is the parameter $B$) are
driven to the infrared attractive fixed points if  the Yukawa
couplings at the GUT scale are large enough.  In the next section we
use the obtained IRQFPs to calculate the masses of some particles of
interest.

Our analysis is constrained by the one-loop RG equations. On the
other hand Abel and Allanach \cite{5} have studied the two-loop
RGEs in the low $\tan\beta$ case.  The difference between one-loop
and two-loop IRQFPs is around 10 per cent.  At the same time the
deviations from the IRQFPs obtained by one-loop RGEs are also
around 10-15 per cent.  However, as we show in the next section
(see Fig.\ref{f3},\ref{f4}), these deviations give negligible
corrections to the Higgs boson masses.  For this reason and in
order to make our arguments simple and clear we follow the
strategy described in Ref.\cite{6} and consider only  one-loop RG
equations to determine the IRQFPs.

\section{Masses of Stops, Sbottoms and Higgs Bosons}

In this section we use the obtained restrictions on the parameter
space coming from the IRQFP behaviour for the computation of the
masses of the Higgs bosons, stops and sbottoms.

We begin with the description of our strategy. As input parameters we
take the known values of the top-quark, bottom-quark and
$\tau$-lepton masses ($m_{t}, m_b, m_{\tau}$), the experimental
values of the gauge couplings \cite{11} $\alpha_3=0.12,
\alpha_2=0.034, \alpha_1=0.017$, the sum of Higgs vev's squared
$v^2=v_1^2+v_2^2=(174.1 GeV)^2$ and the IRQFPs obtained in the
previous section. First, we use the well-known relations between
the running quark and lepton masses and the Higgs v.e.v.s in the
MSSM
\begin{eqnarray} m_t &=& h_t v \sin\beta\,,
\label{mt} \\
m_b &=& h_b v \cos\beta\,, \label{mb} \\
m_{\tau} &=& h_{\tau} v \cos\beta\,. \label{mtau} 
\end{eqnarray}
One can use one of these equations or some combination of them for
 determination of the value of $\tan\beta$. As has been shown in
the previous section the ratio $h_t/h_b$ is almost constant in the
range of possible values of $h_t$ and $h_b$. This opens a
possibility to determine the value of $\tan\beta$, which depends
only on the running masses $m_t$ and $m_{b}$ and very weakly on
the Yukawa couplings $h_t$ and $h_b$. Thus, we use the following
relation for $\tan\beta$:
\begin{equation} \tan\beta=
\frac{m_t}{m_b}\frac{h_b}{h_t}\,.  \label{tan}
\end{equation}
First, we determine $\tan\beta$ using the QCD-corrections to the
running top-quark and bottom-quark masses (see below).  Then, we
apply this value for  calculation of all needed quantities to
determine the SUSY corrections to these masses, which
are again used as input parameters for  evaluation of corrections
to  $\tan\beta$.  After such procedure we are able to obtain a
stable value for $\tan\beta$.

As has been mentioned above, for the evaluation of  $\tan\beta$ we
first need to determine the running top- and bottom-quark masses.
We find them using the  well-known relations to the pole masses
(see e.g.  \cite{4,12,P}). We begin with the top-quark running
mass which is given by the following relation:
\begin{equation}
m_t(m_t)=\frac{m_t^{pole}}{1+ \left(\frac{\Delta
m_t}{m_t}\right)_{QCD} + \left(\frac{\Delta m_t}{m_t}\right)_{SUSY}},
\label{mtpole}
\end{equation}
where $m_t^{pole}=(174.1 \pm 5.4)$ GeV  \cite{13}.
Eq.(\ref{mtpole}) includes the QCD gluon correction (in the
$\overline{MS}$ scheme)
\begin{equation}
\left(\frac{\Delta m_t}{m_t}\right)_{QCD}=\frac{4 \alpha_3}{3 \pi} +
10.92 \left(\frac{\alpha_3}{\pi}\right)^2\,,
\end{equation}
and the stop/gluino correction~\cite{P}
\begin{eqnarray} \nonumber
\left(\frac{\Delta m_t}{m_t}\right)_{SUSY} = &-&
\frac{g_3^2}{12 \pi^2} \Bigl\{B_1(m_t,M_3,\tilde m_{t_1}) +
B_1(m_t,M_3,\tilde m_{t_2})\\
&-& \sin(2\theta_t) \frac{M_3}{m_t} \left[B_0(m_t,M_3,\tilde m_{t_2})-
B_0(m_t,M_3,\tilde m_{t_1})\right] \Bigr\},  \label{msusy}
\end{eqnarray}
where $\theta_t$ is the stop mixing angle, $\tilde m_{t_1}<
\tilde m_{t_2}$, and
\begin{eqnarray}
&&B_n(m_t,m_1,m_2)=-\int_0^1 dx\,x^n \ln\left[\frac{(1-x)m_1^2+x
m_2^2-x(1-x)p^2} {m_t^2} \right].
\end{eqnarray}
We use the following procedure to evaluate the running top mass.
First, we take into account only the QCD correction and find
$m_t(m_t)$ at the first approximation.  This allows us to
determine both the stop masses and the stop mixing angle. Next,
having at hand the stop and gluino masses, we take into account
the stop/gluino corrections. The results are given in the table
below.

Now we consider the bottom-quark running mass  at the $M_Z$ scale,
$m_b(M_Z)$.  The situation is more complicated because the mass of
the bottom $m_b$ is essentially smaller than the scale $M_Z$ and
so we have to take into account the running of this mass from
$m_b$ scale to $M_Z$ scale. The procedure is the following
\cite{P,14,15}: we start with the bottom-quark pole mass,
$m_b^{pole}=4.94 \pm 0.15$ \cite{16}. Then we find the SM
$\overline{MS}$ bottom-quark mass at the $m_b$ scale using the
two-loop $QCD$ corrections
\begin{equation}
m_b(m_b)^{SM}=\frac{m_b^{pole}}{1+
\left( \frac{\Delta m_b}{m_b} \right)_{QCD}}\,,  \label{mbpole}
\end{equation}
where \cite{P,15}
\begin{equation}
\left( \frac{\Delta m_b}{m_b} \right)_{QCD}= \frac{4 \alpha_3(m_b)}{3 \pi} +
12.4 \left( \frac{\alpha_3(m_b)}{\pi} \right)^2\,,
\end{equation}
and $\alpha_3(m_b)$ is the five-flavour three-loop running
$\overline{MS}$ coupling. Then, we evolve this mass to the scale $M_Z$
using a numerical solution of the two-loop (together with
three-loop $O(\alpha_3^3)$) SM RGE's \cite{P,14}. Taking
$\alpha_3(M_Z)=0.12$ one obtains $m_b(M_Z)_{SM}=2.91$GeV. Using
this value we can now calculate sbottom masses and then return
back to take into account the SUSY corrections from massive SUSY
particles
\begin{equation}
m_b(M_Z)=
\frac{m_b(M_Z)^{SM}}{1+ \left( \frac{ \Delta m_b }{ m_b } \right)_{SUSY} }\,.
 \label{mbsusy}
\end{equation}
We approximate these corrections by the sbottom/gluino and stop/chargino
loops (for details see \cite{P} and references therein):
\begin{equation}
\left(\frac{\Delta m_b}{m_b}\right)_{SUSY}=
\left(\frac{\Delta m_b}{m_b}\right)_{\tilde b \tilde g} +
\left(\frac{\Delta m_b}{m_b}\right)_{\tilde t \tilde \chi^{+}}\,.
\end{equation}
The sbottom/gluino contribution is given by (\ref{msusy}) with the
substitution $t \to b$. The sbottom/chargino contribution is as
follows (see \cite{P} for details):
\begin{eqnarray}
\left(\frac{\Delta m_b}{m_b}\right)_{\tilde t \tilde \chi^{+}}&=&
-\frac{h_t^2}{16 \pi^2}\mu
\frac{A_t \tan\beta - \mu}{\tilde m_{t_1}^2-\tilde m_{t_2}^2}
\left[B_0(0, \mu, \tilde m_{t_2}) - B_0(0, \mu, \tilde m_{t_1})\right]
\nonumber \\
&-& \frac{g^2}{16 \pi^2}
\Bigl\{\frac{\mu M_2 \tan\beta}{\mu^2 - M_2^2}
\left[c_t^2 B_0(0, M_2, \tilde m_{t_2}) +
s_t^2 B_0(0, M_2, \tilde m_{t_1})\right] \nonumber \\
&+& (\mu \leftrightarrow M_2) \Bigr\}\,, \label{char}
\end{eqnarray}
where $c_t$ $(s_t)$ is $\cos \theta_t$ $(\sin \theta_t)$.

The running bottom-quark mass slightly depends on the values of
the top-quark Yukawa coupling and SUSY breaking parameters. Hence,
we consider possible restrictions on these parameters. First, from
eq.(\ref{mt}) we have a mathematical condition which requires
$\sin\beta \le 1$. It means that for the value of the running
top-quark mass, $m_t$, we have to take only such values of the
top-quark Yukawa coupling, $h_t$, that give $h_t \ge m_t/v$. This
gives us a restriction from below for the top-quark Yukawa
coupling. One can also impose a restriction from above. In what
follows, we require the values of the top-quark, bottom-quark and
$\tau$-lepton masses to be in  appropriate experimental intervals.
After analyzing  the system of eqs.(\ref{mt}-\ref{mtau}) one can
see that the restriction from above for the Yukawa couplings
follows from the mass of the $\tau$-lepton. This mass is very well
defined experimentally ($m_{\tau}^{pole}=(1.7771 \pm 0.0005)$GeV
\cite{17}) and to get a proper value one imposes the  restriction
on  the values of the Yukawa couplings. Provided these two
conditions are satisfied one has the following intervals for the
third generation of  Yukawa couplings:
 $$h_t \in <0.988, 1.069>, \ \ h_b \in <0.937, 0.982>, \ \
  h_{\tau} \in <0.537, 0.670>\ \ \ \mbox{ for} \ \mu>0,$$
 $$h_t \in <0.924, 0.988>, \ \ h_b \in <0.882, 0.937>, \ \
  h_{\tau} \in <0.475, 0.537>\ \ \ \mbox{for}\ \mu<0.$$
  As one can see, the restriction is larger for the case
of $\mu<0$. These values are obtained through the self-consistent
procedure described above.

Now we can return back and compute the running bottom-quark masses
in each boundary case. It is also  possible  to write down the
corresponding values of  $\tan\beta$ (see the table).
\begin{table}[ht]
\begin{center}
\begin{tabular}{|c|c|c|c|c|c|c|}\hline \hline
$\tan\beta$ & $m_t(M_Z)$ & $m_b(M_Z)$  & $h_t$ & $h_b$ & $m_0^2/m_{1/2}^2$ & $\mu $  \\
\hline 76.4 & 186.1 & 2.24  & 1.069 & 0.982 & 0 & $>0$  \\
  75.5 & 172.0 & 2.16 & 0.988 & 0.937 & 0 & $>0$  \\
  77.3 & 186.2 & 2.21  & 1.069 & 0.982 & 2 & $>0 $ \\
  75.3 & 172.2 & 2.17 & 0.989 & 0.938 & 2 & $>0 $ \\
  44.9 & 172.0 & 3.63  & 0.988 & 0.937 & 0 & $<0$  \\
  46.9 & 160.8 & 3.27 & 0.924 & 0.882 & 0 & $<0 $ \\
  45.3 & 172.1 & 3.60  & 0.989 & 0.938 & 2 & $<0 $ \\
  47.6 & 163.6 & 3.28 & 0.940 & 0.897 & 2 & $<0  $\\ \hline \hline
\end{tabular}
\end{center}
\caption{The values of $\tan\beta$ and running masses of
the top and bottom quarks for both signs of the parameter $\mu$ and
the ratio $m_0^2/m_{1/2}^2=0$ and 2 and for the upper (odd lines)
and lower (even lines) values of top and bottom Yukawa couplings.
}
\end{table}
Due to the stop/gluino, bottom/gluino and stop/chargino  corrections to the
running top and bottom masses the predictions for $\tan\beta$ are different
for different signs of $\mu$. As a consequence, the predictions for the
Higgs bosons masses are also different  in
spite of the fact that these parameters are not explicitly dependent on the
sign of $\mu$ at the tree level.

Now one may  return to eqs.(\ref{mt}-\ref{mtau}) to verify if the
masses obtained  correspond to the upper and lower experimental
bounds. Indeed, there is a  good agreement. In the table we
give the   masses of the top and bottom quarks obtained with the help
of eqs.(\ref{mt}-\ref{mb}). The running masses which are calculated
using  eqs.(\ref{mtpole},\ref{mbpole},\ref{mbsusy}) are
inside the intervals given in the table for different
signs of $\mu$ and the ratio $m_0^2/m_{1/2}^2$.

When calculating the stop and sbottom masses we need to know the
Higgs mixing parameter $\mu$.  For determination of this parameter
we use the well-known relation between the $Z$-boson mass and the
low-energy values of $m_{H_1}^2$ and $m_{H_2}^2$ which comes from
the minimization of the Higgs potential:
\begin{equation}
\frac{M_Z^2}{2}+\mu^2=\frac{m_{H_1}^2+\Sigma_1-
(m_{H_2}^2+\Sigma_2) \tan^2\beta}{\tan^2\beta-1}\,,
\label{MZC}
\end{equation}
where $\Sigma_1$ and $\Sigma_2$ are the one-loop corrections~\cite{18}.
Large contributions to these functions come from stops and
sbottoms.
This equation allows one to obtain the absolute value of $\mu$. As has
already been mentioned, the sign of $\mu$ remains a free parameter.

Having all the relevant parameters at hand we are now able  to
estimate the masses of phenomenologically interesting particles.
With the fixed point type  behaviour we have the only dependence
left, namely on $m_{1/2}$ or the gluino mass, $M_3$. It is
restricted only experimentally: $M_3>144$GeV \cite{11} for
arbitrary values of the squarks masses.

\begin{figure}[t]
     \vspace{-12.5cm}
       \begin{flushleft}
       \leavevmode
       \epsfxsize=16cm
       \epsffile{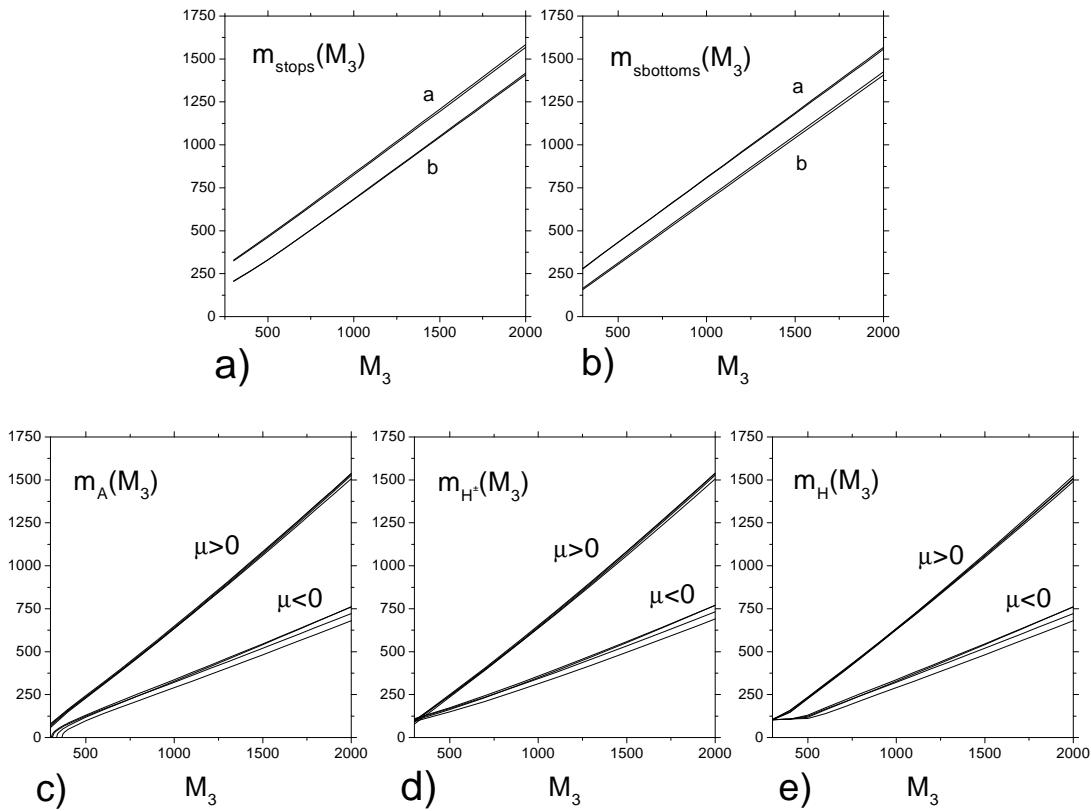}
   \end{flushleft}
\caption{ Masses of the stops a), sbottoms b) and Higgs bosons
c),d),e) as functions of the gluino mass, $M_3$. The curves (a) in
Figs. a) and b) correspond to $\tilde m_{t_2}$ and $\tilde
m_{b_2}$ and the curves (b) correspond to $\tilde m_{t_1}$ and
$\tilde m_{b_1}$ for both signs of $\mu$. Different lines for
masses of the Higgs bosons in both the cases $\mu>0$ and $\mu<0$
are due to the deviations from the IRQFPs and from the uncertainty
in the Yukawa couplings. \label{f3}}
\end{figure}

The stop and sbottom masses are determined by diagonalizing
the corresponding mass matrix \cite{19,20}. The results are shown
in Figs.\ref{f3}a,b. Since the masses are very heavy, the one-loop
corrections to them are not important from the phenomenological
point of view.  For $M_3 \approx 1.5$TeV (which corresponds to
$m_{1/2} \approx 500 $ GeV) we obtain the following numerical
values for the squark masses:
$$\tilde m_{t_1} \approx 1041\ GeV,
\ \tilde m_{t_2} \approx 1195\ GeV, \ \tilde m_{b_1} \approx 1050\
GeV,\ \tilde m_{b_2} \approx 1185\ GeV \ \ \mbox{for} \ \mu>0,$$
$$\tilde m_{t_1} \approx 1050\ GeV, \ \tilde m_{t_2} \approx 1218\
GeV, \ \tilde m_{b_1} \approx 1066\ GeV, \ \tilde m_{t_1} \approx
1196\ GeV \ \ \mbox{for} \ \mu<0.$$
 The dependence   of the squark masses on the allowed variation of the
Yukawa couplings is negligible.

Much interest attracts the mass spectrum of the Higgs
bosons. In the MSSM, after electroweak symmetry breaking, the Higgs
sector consists of five physical states. Their masses obtain large
radiative corrections. In contrast with the low $\tan\beta$ case,
for high $\tan\beta$ they are important for all Higgs bosons and
not only for the lightest one. For the charged bosons and CP-odd
one it is sufficient to take into account the one-loop corrections. As
for the lightest CP-even Higgs boson we also include the two-loop
contributions. For this purpose we need the masses of the stops
and sbottoms calculated above.

\input epsf

The one-loop corrected mass $M_A^2$ is given by the following equation
\cite{18}:
\begin{equation}
m_A^2=m_{H_1}^2 + m_{H_2}^2 + 2 \mu^2 +\Sigma_1 +\Sigma_2 +
\frac{\Delta}{\sin 2 \beta}\,. \label{mA}
\end{equation}

Following our strategy and using the IRQFPs obtained above we have
the value of $m_A$ which depends only on the gluino mass $M_3$
shown in Fig.\ref{f3}c.  One can  see that the requirement of
positivity of $m_A^2$ excludes the region with small $M_3$. This
restriction is important when analyzing  the lightest Higgs boson
mass.
 In the most promising region  $M_3>1$ TeV ($m_{1/2}>300$ GeV),
 however, for the both cases $\mu>0$ and $\mu<0$
the mass of the CP-odd Higgs boson is too heavy to be detected in
the near future
$$m_A>1100 \ \mbox{GeV  for} \ \  \mu>0, \ \ \
m_A>570 \ \mbox{GeV for}\ \  \mu<0,$$
when $M_3 \approx 1.5$TeV.

The mass of the charged Higgs bosons $m_{H^{\pm}}^2$ with one-loop
corrections can be written as follows \cite{21}:
\begin{equation}
m_{H^{\pm}}^2 = m_A^2 + M_W^2 + \Delta_c^2\,. \label{mHc} \\
\end{equation}
In  Fig.\ref{f3}d one can see its behaviour for different signs
of $\mu$. Like in the case of low $\tan\beta$ the charged Higgses
are heavy but there is a big difference between the cases with $\mu>0$
and $\mu<0$. One can find the following values at the typical
scale $M_3 \approx 1.5$TeV:
$$m_{H^{\pm}} > 1105 \ \mbox{GeV for}
\ \  \mu>0, \ \ \   m_{H^{\pm}}> 575\ \mbox{GeV for}\ \  \mu<0.$$

The most phenomenologically  interesting particle is the lightest
Higgs boson $h$. The two-loop corrected masses of the CP-even
Higgs bosons, $h$ and $H$ are given by
\begin{equation}
m_{H,h}^{2} = \frac 12 (Tr(M^2) \pm \sqrt{(Tr(M^2))^2-4\det(M^2))}\,,
\label{h}
\end{equation}
where $M^2$ is the $2\times2$ symmetric mass matrix  given in
Ref.~\cite{22}. One can see in Fig.\ref{f3}e that
 the heaviest  CP-even Higgs boson, $H$, is too heavy for
$\mu>0$ and not so heavy for $\mu<0$ but still far away from the
range of the nearest experiments. For the typical value of the
gluino mass $M_3 \approx 1.5$ TeV
 its mass is
$$m_H > 1100 \ \mbox{GeV for} \ \  \mu>0, \ \ \ m_H > 570 \ \mbox{GeV  for}
 \ \ \mu<0.$$
 The dependence on the deviations from the IRQFPs is not
relevant
as one can see in Fig.\,\ref{f3}e, and $H$-boson is
too heavy to be detected in the near future.

\begin{figure}[t]
     \vspace{-15.5cm}
       \begin{flushleft}
       \leavevmode
       \epsfxsize=16cm
       \epsffile{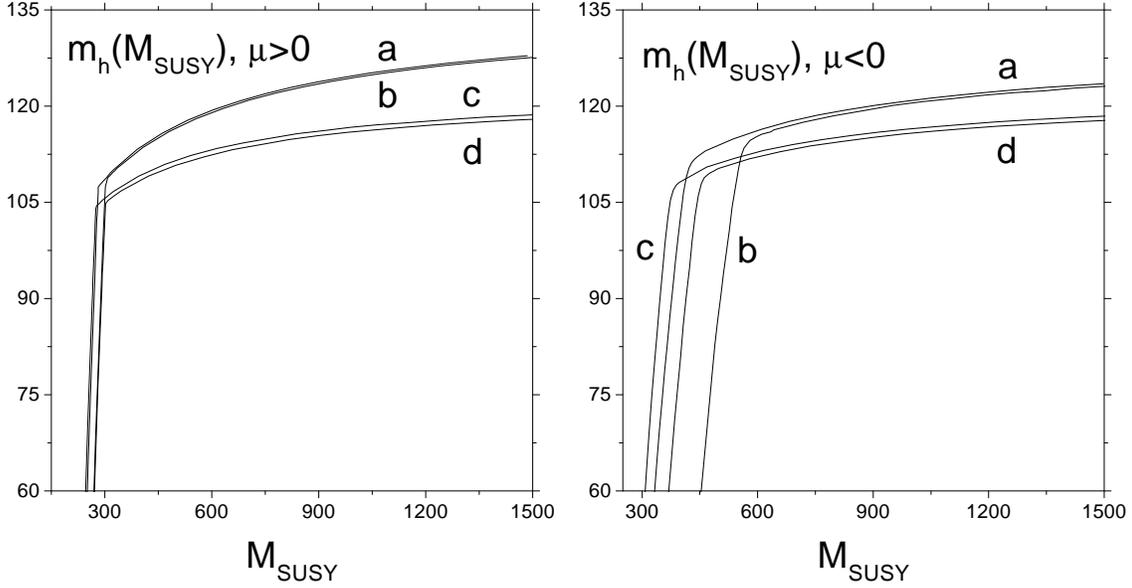}
   \end{flushleft}
\caption{The mass of the lightest Higgs boson, $h$, as a function
of $M_{SUSY}$ for different signs of $\mu$. The curves (a,b)
correspond to the upper limit for the Yukawa couplings and
$m_0^2/m_{1/2}^2=0$ (a) and $m_0^2/m_{1/2}^2=2$ (b). The curves
(c,d) correspond to the lower limit for the Yukawa couplings and
$m_0^2/m_{1/2}^2=0$ (c) and $m_0^2/m_{1/2}^2=2$ (d). The allowed
values of the lightest Higgs boson mass lie inside the areas
marked by these lines. \label{f4}}
\end{figure}

The situation is different for the lightest Higgs boson, $h$,
which is much lighter. As one can see from  Fig.\ref{f4} the
deviations of the Yukawa couplings from their IRQFPs give rather
wide interval of values for  $m_h$ in both the cases $\mu>0$ and
$\mu<0$.
One has  the following values of  $m_h$ at a typical scale
$M_3=1.5$TeV
\begin{eqnarray}
m_h&=&129.3-0.1-7.2 \pm 5  \ \mbox{GeV,  for} \   \mu>0 \,, \nonumber \\
m_h&=&121.8-0.3-4.1 \pm 5 \ \mbox{GeV, for} \   \mu<0 \,.  \nonumber
\end{eqnarray}
The first uncertainty is connected with the deviations from the
IRQFPs for mass parameters, the second one with the Yukawa
coupling IRQFPs, and the third one is due to the experimental
uncertainty in the top-quark mass. One can immediately see that
the deviations from the IRQFPs for mass parameters are negligible
and only influence the steep fall of the function on the left,
which is related to the restriction on the CP-odd Higgs boson mass
$m_A$.  In contrast with the low $\tan\beta$ case, where the
dependence on the deviations from the fixed points was about $1$
GeV, in the present case the dependence is much stronger. The
experimental uncertainty in the strong coupling constant
$\alpha_s$ is not included  because it is negligable compared to
those of the top-quark mass and the Yukawa couplings and is not
important here contrary to the low $\tan\beta$ case (see
Ref.~\cite{6}).

We show in Fig.\ref{f4} the mass of the lightest Higgs boson
 as a function of  $(\tilde m_{t_1} \tilde
m_{t_2})^{1/2}$ which is  usually referred to as the $M_{SUSY}$ scale.
In this case we have the following values for the lightest Higgs boson
for a typical value $M_{SUSY}=1$ TeV ($M_3 \approx 1.3$TeV):
\begin{eqnarray}
m_h&=&128.2 -0.4 - 7.1 \pm 5 \ \mbox{GeV,  for} \  \mu>0 \,, \nonumber \\
m_h&=&120.6 -0.1 - 3.8 \pm 5 \ \mbox{GeV, for} \  \mu<0 \,.  \nonumber
\end{eqnarray}

One can see that for large $\tan\beta$ scenario the masses of the
lightest  Higgs boson are typically around 120 GeV which is too
heavy for observation at LEP II.

\input epsf

\section{Summary and Conclusion}

Thus, we have analyzed the fixed point behaviour of SUSY breaking
parameters in the large $\tan{\beta}$ regime. We made this
analysis assuming that the Yukawa couplings are initially large
enough to be driven at infrared scales to their Hill-type quasi
fixed points. This  corresponds to a  possible $SO(10)$ Grand
Unification scenario with radiative EWSB.

We have found  that solutions of RGE's for some of the SUSY
breaking parameters become insensitive to their initial values at
unification scale. This is because at infrared scale they are
driven to their IR quasi fixed points. These fixed points are used
to make predictions for the masses of the Higgs bosons, stops and
sbottoms. We have taken into account possible deviations from
quasi IR fixed points. This leads to uncertainties in mass
predictions which are much bigger than in the low $\tan\beta$
case, however, still dominated by the top mass experimental
bounds.

In the IRQFP scenario all the Higgs bosons except for the lightest
one are found to be too heavy to be accessible in the nearest
experiments. The same is true for the stops and sbottoms. From
this point of view the situation is the same as in the low
$\tan\beta$ case and essentially coincides with the results of
more sophisticated analyses \cite{18}. The lightest neutral Higgs
boson, contrary to the  low $\tan\beta$ case, is not within the
reach of LEPII leaving hopes for the Tevatron and LHC.
 \vglue 0.5cm

{\bf Acknowledgments}

\vglue 0.5cm

We are grateful to W. de Boer and G. K. Yeghiyan
for useful discussions.  Financial support from RFBR grants \#
98-02-17453 and 96-15-96030 is kindly acknowledged.

\end{document}